 \documentclass[twocolumn,twoside,showpacs,preprintnumbers,superscriptaddress,amsmath,amssymb,prl]{revtex4}
 
 \usepackage{cases}
 \usepackage{CJK}
 \usepackage{txfonts}
 \usepackage{mathrsfs}
 \usepackage{amssymb}
 \usepackage[dvips]{graphicx}
 \usepackage{dcolumn}
 \usepackage{bm}
 \usepackage{subfigure}
 \usepackage{floatflt}
 \usepackage{float}
 \usepackage{rotating}
 \usepackage{multirow}
 \usepackage[bookmarksnumbered,bookmarksopen,colorlinks,citecolor=blue,linkcolor=blue]{hyperref} %
 \usepackage{booktabs}
 \usepackage{epsfig,graphics,psfrag,amsmath}
 \usepackage{sidecap}

\usepackage{dcolumn}
\usepackage{bm}
\def\thu{Department of Physics, Tsinghua University, Beijing 100084, China}

\def\imp{Institute of Modern Physics, Chinese Academy of Sciences, Lanzhou 730000, China}
\def\hnu{Institute of Particle and Nuclear Physics, Henan Normal University, Xinxiang 453007, China}

\def\cicq{Center of High Energy Physics, Tsinghua University, Beijing 100084, China}
\def\hnas{Institute of Nuclear Science and Technology, Henan Academy of Sciences, Zhengzhou, 450015, China}

\newcommand{\scattering}{18.9^{+1.3}_{-1.2}}
\newcommand{\effective}{1.9^{+1.3}_{-1.0}}
\newcommand{\Raverage}{4.12 \pm 0.12}
\newcommand{\RHigh}{2.8 \pm 0.1}
\newcommand{\RLow}{4.9 \pm 0.2}
\begin{document}
\preprint{APS/123-QED}

\title {Extract neutron-neutron  interaction strength and  spatial-temporal dynamics of neutron emission from two-particle correlation function}

\author{Dawei Si}\email{sdw21@mails.tsinghua.edu.cn}
 \affiliation{\thu}

 \author{Sheng Xiao}  
 \affiliation{\thu}

\author{Zhi Qin}
 \affiliation{\thu}
 
\author{Yuhao Qin}
 \affiliation{\thu}

\author{Junhuai Xu}
 \affiliation{\thu}

\author{Baiting Tian}
 \affiliation{\thu}

\author{Boyuan Zhang}
 \affiliation{\thu}

\author{Haojie Zhang}
 \affiliation{\thu}
 
\author{Dong Guo}
 \affiliation{\thu}

\author{Yijie Wang}\email{yj-wang15@tsinghua.org.cn}
 \affiliation{\thu}

\author{Xiaobao Wei}
 \affiliation{\hnu}

\author{Yibo Hao}
 \affiliation{\hnu}
 
\author{Zengxiang Wang}
 \affiliation{\hnu}
 
 \author{Tianren Zhuo}
 \affiliation{\hnu}
 
\author{Chunwang Ma}
 \affiliation{\hnu}
\affiliation{\hnas}

\author{Yuansheng Yang}
 \affiliation{\imp}

\author{Xianglun Wei}
 \affiliation{\imp}

\author{Herun Yang}
 \affiliation{\imp}

\author{Peng Ma}
 \affiliation{\imp}

\author{Limin Duan}
 \affiliation{\imp}

\author{Fangfang Duan}
 \affiliation{\imp}
 
 \author{Kang Wang}
 \affiliation{\imp}

\author{Junbing Ma}
 \affiliation{\imp}

\author{Shiwei Xu}
 \affiliation{\imp}

\author{Zhen Bai}
 \affiliation{\imp}

\author{Guo Yang}
 \affiliation{\imp}

\author{Yanyun Yang}
 \affiliation{\imp}

\author{Zhigang Xiao}\email{xiaozg@tsinghua.edu.cn}
 \affiliation{\thu}
 \affiliation{\cicq}

\date{\today}

\begin{abstract}
 The neutron-neutron ($nn$) correlation function has been measured in 25 MeV/u $^{124}$Sn+$^{124}$Sn reactions. 
 Using the  Lednick\'y-Lyuboshitz  approach, the $nn$ scattering length and effective range ($f_{0}^{nn}$, $d_{0}^{nn}$), as well as  the reduced space-time size $R^{(0)}$  of the neutron emission source are simultaneously extracted as ($\scattering$ fm, $\effective$ fm) and $\Raverage$ fm, respectively.  The measured $nn$ scattering length is consistent with the results obtained in the low-energy scattering  $^{2}{\rm H}(\pi^{-},\gamma)2n$, indicating heavy-ion collisions 
can serve as an effective  approach for measuring $nn$ interactions and further investigating the charge symmetry breaking of nuclear force. The space-time size extracted from  momentum-gated correlation functions exhibits clear dependence on the pair momentum,  with $R^{(0)}=\RHigh $ fm and  $\RLow$ fm being determined for the high and low momentum neutrons, respectively. 

\end{abstract}

\maketitle

{\it Introduction - } The neutron-neutron ($nn$) strong interaction is of fundamental significance to understand the charge symmetry breaking (CSB) of nuclear force \cite{CSB1,CSB2}. However, direct measurement of $nn$ scattering length($f_{0}^{nn}$) using free neutrons is not feasible because of the unavailability of neutron target. Instead,  the $nn$ scattering length  $f_{0}^{nn}$ has been measured indirectly in the few-body reactions with two neutrons in the final states, like $^{2}{\rm H}(n,p)2n$, $^{2}{\rm H}(\pi^{-},\gamma)2n$ and $^{3}{\rm H}(t,\alpha)2n$ \cite{n-d1,pi-d1,3He-t}. The discrepancy of $f_{0}^{nn}$ from these channels exists, indicating the three-body force is at work, and calling for further measurements.

Alternatively, it was first proposed by  Lednick\'y and Lyuboshitz (LL) to constrain the $nn$ interaction from  the correlation function (CF) of neutron pair  with low relative momentum \cite{LL1}, a method deduced from the generalization of the intensity interferometry invented by Hanbury-Brown and Twiss (HBT) \cite{HBT1,HBT2}.  Since the proton-proton ($pp$) CF in heavy ion reactions was formulated \cite{p_pbar},  the application of the LL method has been extensively employed to study the final-state interactions of various baryon pairs,  such as $\bar p \bar p$, $p\Lambda$, $p\Omega^{-}$ and $\Lambda\Lambda$ pairs \cite{p_pbar,HIGHCF1,HIGHCF2,HIGHCF3,Theory1}. Yet, the extraction of $nn$ scattering length from CF has not been reported since the LL approach was formulated.

The CF method can also be utilized to extract the spatial-temporal characteristics of the emission source \cite{Wang:2021mrv,Theory2,Xu:2024dnd}. Enormous experimental analysis have been performed using the probes of pions and kaons to infer the spatial feature of the source formed in the relativistic and ultrarelativistic energy experiments \cite{meson1,meson2,meson3,meson4,meson5}. Furthermore, the space information encoded in the CF has also be utilized to study the valence nucleon distribution of unstable nuclei \cite{nucl_structure1,nucl_structure2,nucl_structure3}. More interestingly, going down to intermediate and low energy heavy-ion reactions, the emission time scale, which is much longer, brings  significant effect to the final CF   \cite{Spacetime1,Wang:2021mrv,Xiao:2006fdi}. And hence the CF method applied to $pp$ and $nn$ offers the opportunity to differentiate the emission time difference between neutrons and protons, which may provide a new probe to constrain the stiffness of nuclear  symmetry energy under hot debate currently \cite{Spacetime2,Spacetime3}.

Although experimental $pp$ CFs are more copious, the $nn$ CFs are very scarce. So far there are mainly two  heavy-ion experiments at intermediate and low energies reporting the $nn$ CFs measured over 20 years ago.  Colonna et al. first observed the difference between the transverse and longitudinal $nn$ CFs in $^{18}$O + $^{26}$Mg at 130 MeV 
 \cite{nn_cor1}. The CHIC Collaboration  measured the $nn$ CF in 45 MeV/u $^{58}$Ni + $^{27}$Al, $^{nat}$Ni and $^{197}$Au reactions, and observed the increasing correlation strength  with the cut on the total-momentum of neutron pairs 
 \cite{nn_cor3,nn_cor4,nn_cor5}. In these analyses the emission source size and the average emission time scale were determined by fixing the $nn$ interaction parameters  \cite{nn_cor6}.

In this letter, the $nn$ CF in the neutron rich $^{124}$Sn+$^{124}$Sn reactions at 25 MeV/u is reported. A novel method to subtract the cross talk effect is introduced to avoid the significant distortion on the CF due to very strong cut. The $nn$ scattering length $f_{0}^{nn}$  and the effective range $d_{0}^{nn}$, as well as the reduced emission size $R^{(0)}$ are simultaneously extracted based on LL fit. The ($f_{0}^{nn}$, $d_{0}^{nn}$) are compared to the results extracted from scattering experiments, and the space-time feature of the neutron emission source is discussed.     

\begin{figure}[!htb]
\includegraphics[width=0.85\hsize]{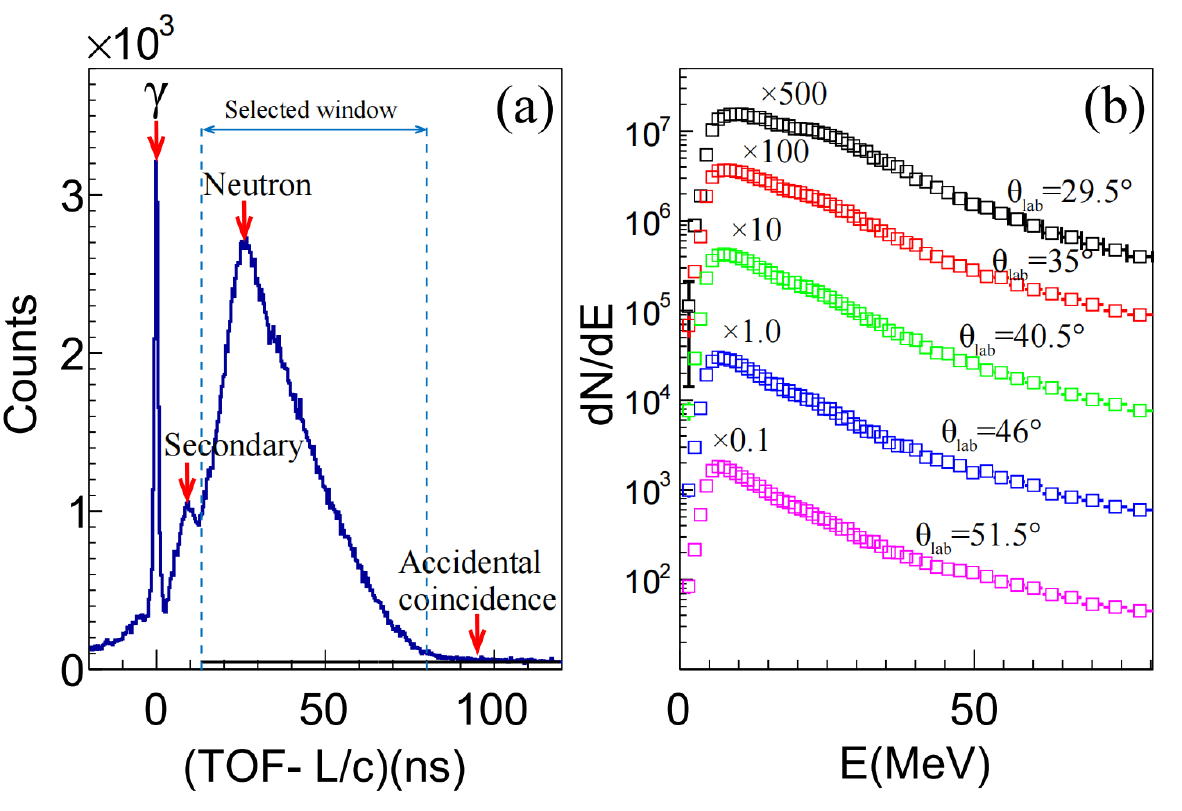}
\caption{(Color online) (a)The experimental TOF distribution. The $\gamma$ peak is used to calibrate the zero time.  (b) The kinetic energy spectra of neutrons at various laboratory  angles.}
\label{fig1}
\end{figure}

{\it Experiment -} The experiment was performed at the final focal plane of the radioactive ion beam line at Lanzhou (RIBLL1)  with the compact spectrometer in heavy ion experiment (CSHINE) \cite{CSHINE1,CSHINE2,CSHINE3,CSHINE4}. The $^{124}$Sn beam was delivered by the cyclotron of the heavy
ion research facility at Lanzhou (HIRFL), bombarding on a $^{124}$Sn target with a thickness of 1.0 $\rm mg/cm^{2}$. Neutrons were detected by a neutron array consisting of twenty $15  \times  15  \times  \rm 15 cm^3$ plastic scintillators coupled with a 2-inch  photomultiplier tube, which are installed in 5 columns and 4 rows on the spherical surface with a distance of $L=200$~cm  to the target, covering the laboratory polar angle  $27^{\circ}< \theta_{\rm lab} <53^{\circ}$. 
The relative angle between units varies from $6^{\circ}$ to $30^{\circ}$, ensuring a sufficient coverage of relative momentum range for CF measurement.
For more details of the neutron array, one can refer to \cite{CSHINE5}. The energy of neutron was measured with the time of flight (TOF) method, using  $\rm BaF_{2}$ as the start timing detector. 
The overall time resolution in the beam experiment is 0.74 ns, resulting from the inherent time resolution  (0.2 ns) and the variation of the neutron interaction position in the scintillator (0.7 ns).
Fig. \ref{fig1} (a) illustrates the experimental TOF distribution. Besides the neutrons and the $\gamma$-rays from the triggered events, there also exist secondary and accidental coincidence backgrounds. Fig.  \ref{fig1} (b) presents the neutron energy spectra at different angles $\theta_{\rm lab}$. The inverse slope parameters of the spectra vary from $13.0\pm0.1$ to  $22.9 \pm 0.1$ MeV depending on $\theta_{\rm lab}$ \cite{CSHINE5}. To reduce the influence of background, the events within the TOF range from 20 ns to 87.5 ns, corresponding to the neutron kinetic energy of 3 MeV to 76 MeV, are selected to generate the $nn$ CF.

The CF was constructed from the relative momentum $k^{*}=\left|\mathbf{p}_{1}-\mathbf{p}_{2}\right|/2$ of the two particles in the pair rest frame (PRF), 
where $\mathbf{p}_{1}$ and $\mathbf{p}_{2}$ are the momenta of the two particles. The
experimental CF is defined as
\begin {equation}
C_{\rm exp}(k^{*})=A\frac{N^{\rm same}(k^{*})}{N^{\rm mix}(k^{*})},
\label{eq1}
\end {equation}
where $N^{\rm same}(k^{*})$ is the relative momentum  distribution with two particles coming from the same event, and $N^{\rm mix}(k^{*})$  is the reference distribution generated by event mixing. The normalization parameter $A$ is determined by requiring  that $C_{\rm exp}(k^{*}) = 1$ at large relative momenta ($k^{*} > 40$ MeV/c). 

\begin{figure}[!htb]
\includegraphics[width=0.85\hsize]{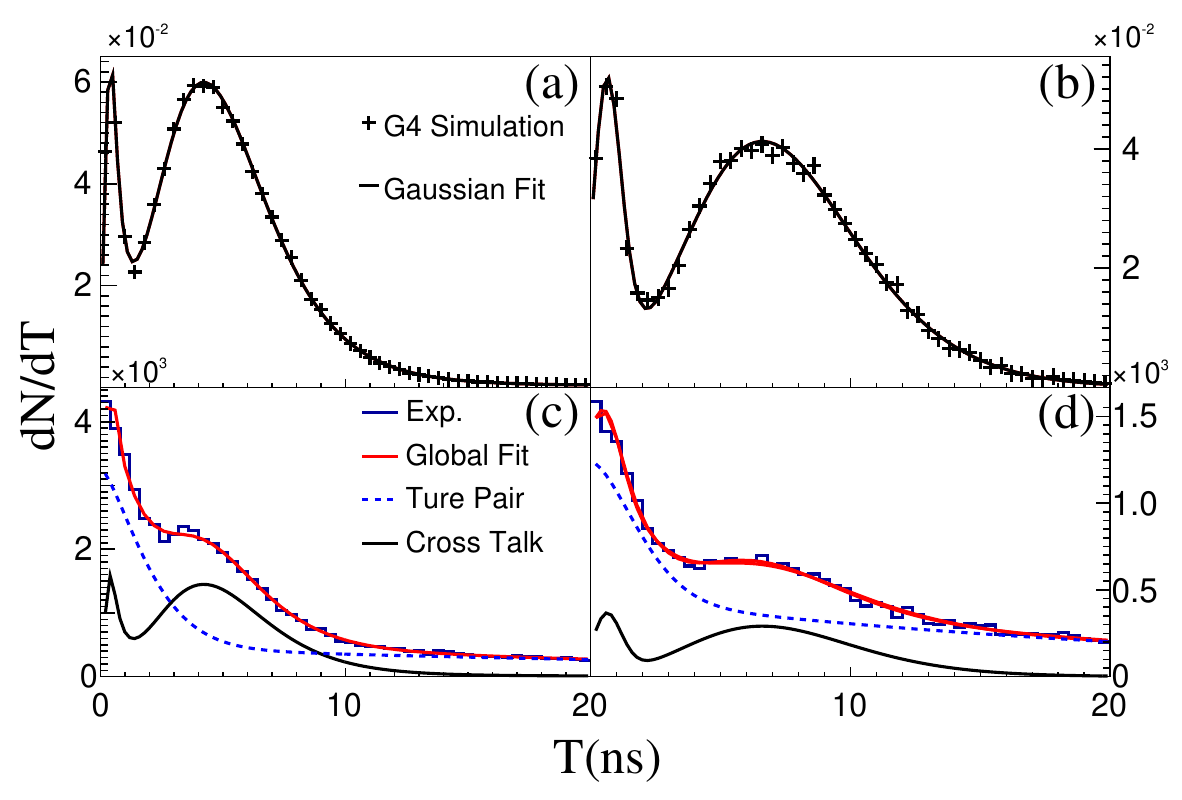}
\caption{(Color online) The TOF difference ($\Delta T$) distributions between two fired detectors for adjacent (a,c) and diagonal (b,d) configurations. Panels (a) and (b) present the cross talk events in Geant4 simulations (cross), fitted with a double asymmetric Gaussian function (black curve). Panels (c) and (d) present the experimental distribution (histogram) decomposed to the cross talk events (black curve) and true two-neutron events (dashed curve).}
\label{fig2}
\end{figure}

The most important effect to be corrected is the cross talk. It arises from a single neutron scattering within the neutron array, and forms the pseudo two-neutron events which are encompassed in the measured raw CF $C_{\rm exp}(k^{*})$.
Additionally, a background component of accidental coincidences is also convoluted, which exhibits a uniform distribution in the TOF spectrum \cite{CSHINE5}. Hence, the genuine $nn$ CF $C_{\rm nn}(k^{*})$ can be obtained by subtracting these two additional components from the experimental CF as
\begin {equation}
\begin {split}
&C_{\rm nn}(k^{*})=A_{nn}\frac{N^{\rm same}_{\rm nn}(k^{*})}{N^{\rm mix}_{\rm exp}(k^{*})}\\
&=\frac{A_{\rm nn}}{1-\lambda_{\rm ct}-\lambda_{\rm ac}}\left(\frac{C_{\rm exp}(k^{*})}{A_{\rm exp}}-\lambda_{\rm ct}\frac{N^{\rm mix}_{\rm ct}(k^{*})}{N^{\rm mix}_{\rm exp}(k^{*})}-\lambda_{\rm ac}\frac{N^{\rm same}_{\rm ac}(k^{*})}{N^{\rm mix}_{\rm exp}(k^{*})}\right),
\end {split}
\label{eq2}
\end {equation}
where $A_{\rm nn}$ and $A_{\rm exp}$ refer to the normalization parameters of the genuine $nn$ CF and the measured raw CF, respectively.  $N^{\rm same}_{\rm ct}(k^{*})$ and $N^{\rm same}_{\rm ac}(k^{*})$ are the relative momentum distributions of cross talk events and accidental coincidence events, and $\lambda_{\rm ct}$ and $\lambda_{\rm ac}$ represent the corresponding proportions, respectively. $\lambda_{\rm ac}$ is obtained by fitting the TOF spectrum \cite{CSHINE5}. And $\lambda_{\rm ct}$ can be derived from the TOF difference spectrum of experimental two-body events, as illustrated below.

To correct the cross talk effect, we consider  two primary configurations, namely the secondary particle produced by the single incident neutron fires the adjacent unit (adjacent configuration) or fires the diagonally neighbouring unit (diagonal configuration).  
Figure \ref{fig2} (a-b) presents the TOF difference between the two fired units in Geant4 simulations for adjacent configuration (a) and diagonal configuration (b), respectively. Two separate peaks are visible. The left narrow one is associated with secondary $\gamma$ rays while the wider peak corresponds to the scattered neutrons featuring a longer flight time. In the diagonal configuration (b), the neutron peak  shifts further rightward due to the longer flight distance. The distribution can be described by a double asymmetric Gaussian function. 

 In the experiment, since the neutron detectors are placed equidistantly in both $x$ and $y$ direction, the flight time of the secondary particles generated by cross talk is relatively fixed, resulting in peak structures in the time difference spectra of two-body events for both adjacent and diagonal configurations, as  shown in Fig. \ref{fig2} (c) and (d). The double peaked structure of the $\Delta T$ spectra can be used to quantify the contribution of cross talk effect. Assuming the cross talk distribution keeps the shape obtained by Geant4 simulations, and using the exponential distribution to describe the true neutron pair time difference, one can decompose the experimental TOF difference spectra with $\lambda_{\rm ct}$ being a fitting parameter. By employing this method, one can derive $\lambda_{\rm ct}$  and avoid using the strict rejection strategy which may bring large distortion to the two-neutron CF \cite{Ghetti1}.

\begin{figure}[!htb]
\includegraphics[width=0.85\hsize]{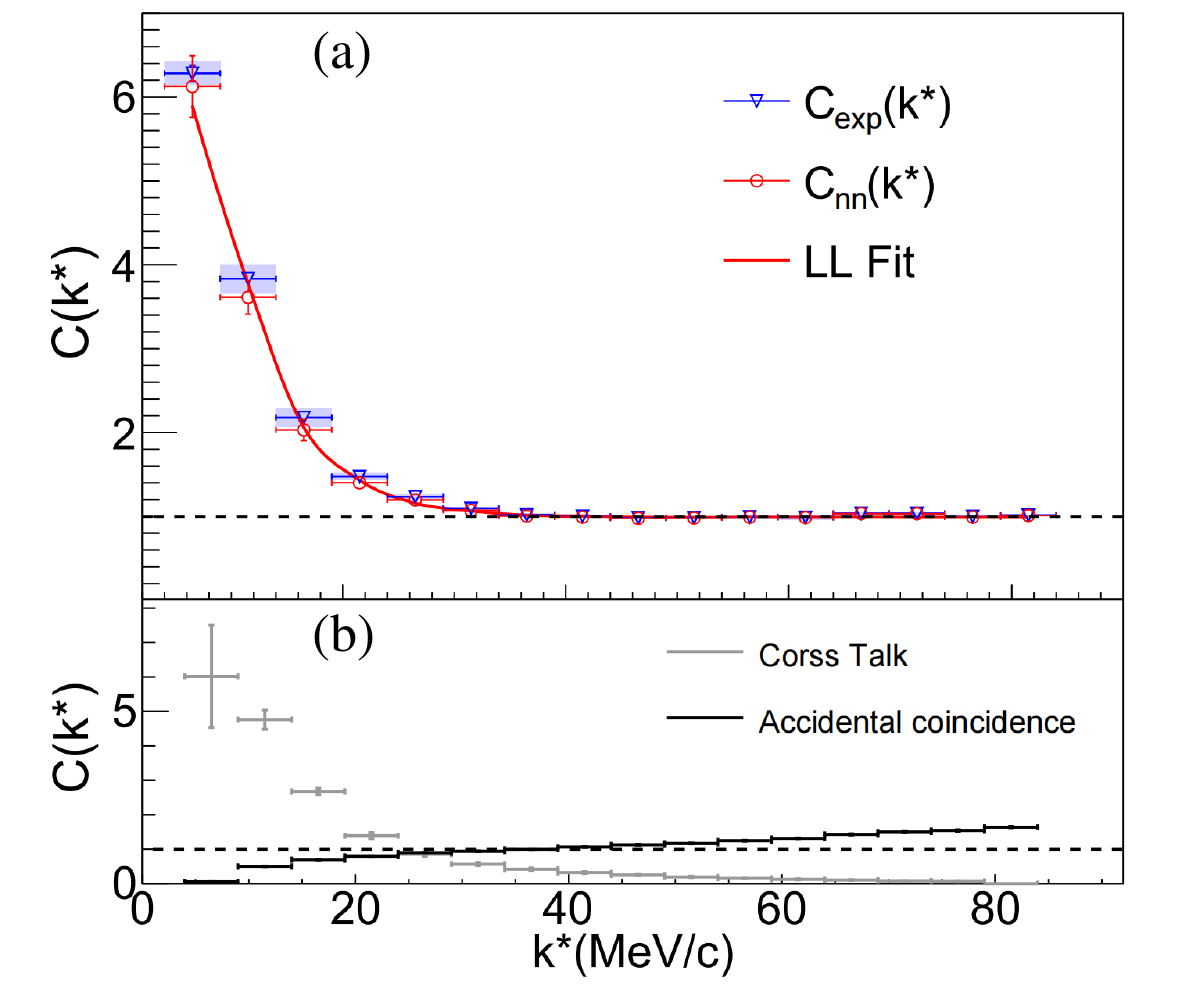}
\caption{(Color online) (a) The raw $nn$ CF $C_{\rm exp}(k^{*})$(blue triangle) and the genuine (red circles) $nn$ CF $C_{\rm nn}(k^{*})$ corrected by Eq. (\ref{eq2}). The blue squares denote the uncertainty of flight distance. The  red solid curve denotes the fitting results of LL model.  (b)  the relative momentum distributions of the cross talk events and the accidental coincidence background.}
\label{fig3}
\end{figure}

{\it Results -} Fig. \ref{fig3} presents the experimental $nn$ CF. The measured raw CF $C_{\rm exp}(k^{*})$ obtained by Eq. (\ref{eq1}) is shown by the blue triangles. The systematic uncertainties, depicted by  the squares, stem from the uncertainty of the flight distance due to the variation of the interaction position within the scintillator. The genuine $nn$ CF $C_{\rm nn}(k^{*})$ is presented by the red circles after subtracting the cross talk and the accidental coincidence using Eq. (\ref{eq2}). Fig. \ref{fig3} (b) presents the relative momentum distributions of cross talk and accidental coincidence events, respectively. The cross talk distribution are significantly enhanced at small relative momenta, resulting in an overestimation of the CF, while the accidental coincidence distribution decreases at small relative momenta and cancels partly the cross talk effect. For the genuine $nn$ CF in panel (a), the uncertainty was calculated bin by bin according to Eq. (\ref{eq2}), including the uncertainty of experimental statistics, flight distance, Monte Carlo simulation in panel (b) and the fitting uncertainty of $\lambda_{\rm ct}$. 

To extract the $nn$ interaction strength and the source distribution, the $nn$ CF is analyzed  using the LL approach \cite{nucl_structure2,LL1,LL2}. Theoretically the CF can be expressed as

\begin {equation}
C(k^{*})=\int S(\mathbf{r},t,\mathbf{p})\left|\Phi(\mathbf{k^{*}},\mathbf{r})\right|^{2}d^{3}r, 
\label{eq3}
\end {equation}
where $\mathbf{r}$ is the relative distance between the particle pair, $S(\mathbf{r},t,\mathbf{p})$ is the distribution of this relative distance for particles emitted in the collision, which is sampled by the single particle emission function parameterized as
\begin {equation}
g(\mathbf{r},t,\mathbf{p})\propto \exp(-r^{2}/R^{2}-t^{2}/\tau^{2})Y(\mathbf{p}) .
\label{eq4}
\end {equation}

Here R and $\tau$ are two parameters characterizing the spatial and temporal feature of the source. The variables  $\mathbf{r}, \mathbf{p}$ and $t$ in Eq. (\ref{eq4}) refer to the rest frame of the source. To incorporate the resolution and detection efficiency of the apparatus, the neutron momenta were selected by randomly sampling the experimental
yield $Y(\mathbf{p})$. Phase-space points generated in the rest frame of the source were Lorentz boosted into the PRF. The relative wave function $\Phi(\mathbf{k^{*}},\mathbf{r})$ is obtained by the effective range expansion \cite{LL1,expansion_range}.

The fitting curve using LL model reproduces well the corrected $nn$ CF, as plotted in Fig. \ref{fig3} (a). 

\begin{figure}[!htb]
\includegraphics[width=0.85\hsize]{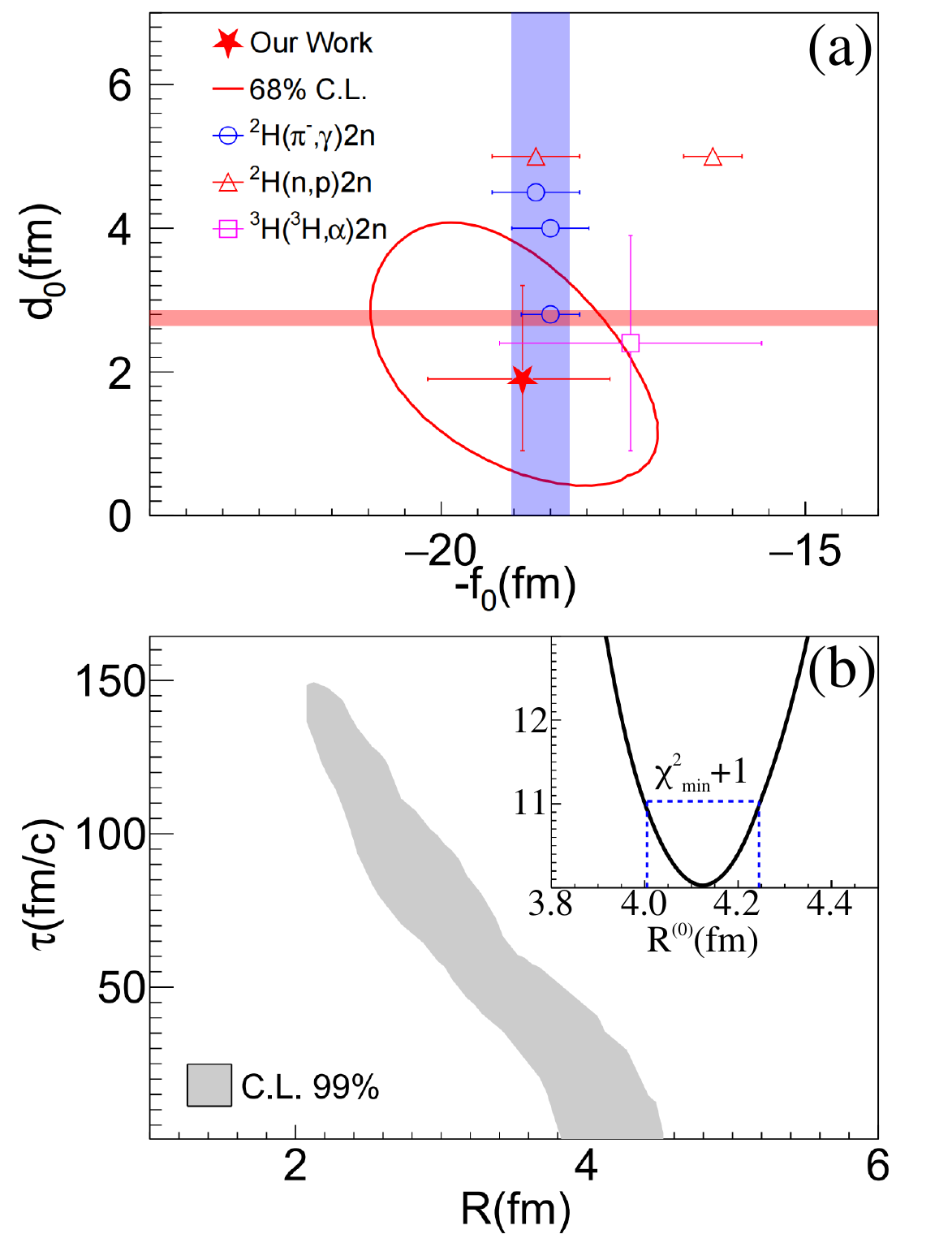}
\caption{(Color online) (a) the fit result of $nn$ interaction parameters $f_{0}^{nn}, d_{0}^{nn}$ from $nn$ CF in comparison to the results from scattering experiments. (b) the ($R$, $\tau$) contours at 99\% confidence levels. The inset illustrates the variation of the minimum $\chi^2$ as a function of the reduced source sizes $R^{(0)}$ at $\tau=0$.}
\label{fig4}
\end{figure}

Now the $nn$ interaction parameters ($f_{0}^{nn},d_{0}^{nn}$) and the  source parameters ($R,\tau$)  can be extracted by LL model \cite{LL4}. In the Fermi energy region,  both the space and time extension influence the CF. To avoid the impact of the space-time ambiguity \cite{ambiguity1} on the interaction parameters, we first fix $\tau = 0$. The best fit of $f_{0}^{nn}=\scattering ~{\rm fm}$ and $d_{0}^{nn}=\effective ~{\rm fm}$ are obtained and plotted in Fig. \ref{fig4} (a) with the combined 68\% confidence level contour. For comparison, the $f_{0}^{nn}$ measured by reaction $^{2}{\rm H}(n,p)2n$ \cite{n-d1,n-d2}, $^{2}{\rm H}(\pi^{-},\gamma)2n$ \cite{pi-d1,pi-d2,pi-d3} and $^{3}{\rm H}(t,\alpha)2n$ \cite{3He-t} are also shown. The blue band represents the widely adopted value of $f_{0}^{nn}$ \cite{f0band}, while the red band indicates the recommended value of  $d_{0}^{nn}$ \cite{d0band}. Remarkably, the measured ($f_{0}^{nn},d_{0}^{nn}$)  from the current CF method is consistent with previous determinations of low-energy scattering experiment, demonstrating  the validity of the LL method  to study $nn$ interactions by measuring the  CF in heavy ion reactions. Worth mentioning, since the uncertainty is comparable to the CSB amplitude, more precise measurements are required definitely in the future to address the CSB using  CF method.  
 
The space-time size of the neutron emission source is simultaneously inferred.  As shown in the inset of Fig. \ref{fig4} (b), the reduced space-time volume (at $\tau =0$) is also obtained to be $ R^{(0)}= \Raverage$ fm, which should be considered as the upper limit for the actual source sizes and is comparable to the results obtained under similar experimental conditions \cite{st_volume1,st_volume2,st_volume3}.  To see further the space-time correlation, we release the condition of $\tau=0$ in the LL  fit. Fig. \ref{fig4} (b) presents the contour plots of emission source parameters ($R$,$\tau$) under 99\% confidence level. It can be seen that $R$ and $\tau$  are mutually coupled and the slope of the distribution is determined by the average relative velocity of the particle pair. On 99\%  confidence level, the emission time scale varies in the range of $0-150$ fm/c. The $R\text-\tau$ distribution reveals the space-time ambiguity of the angle-averaged CF in low energy heavy ion collisions  \cite{ambiguity2}.

\begin{figure}[!htb]
\includegraphics[width=0.85\hsize]{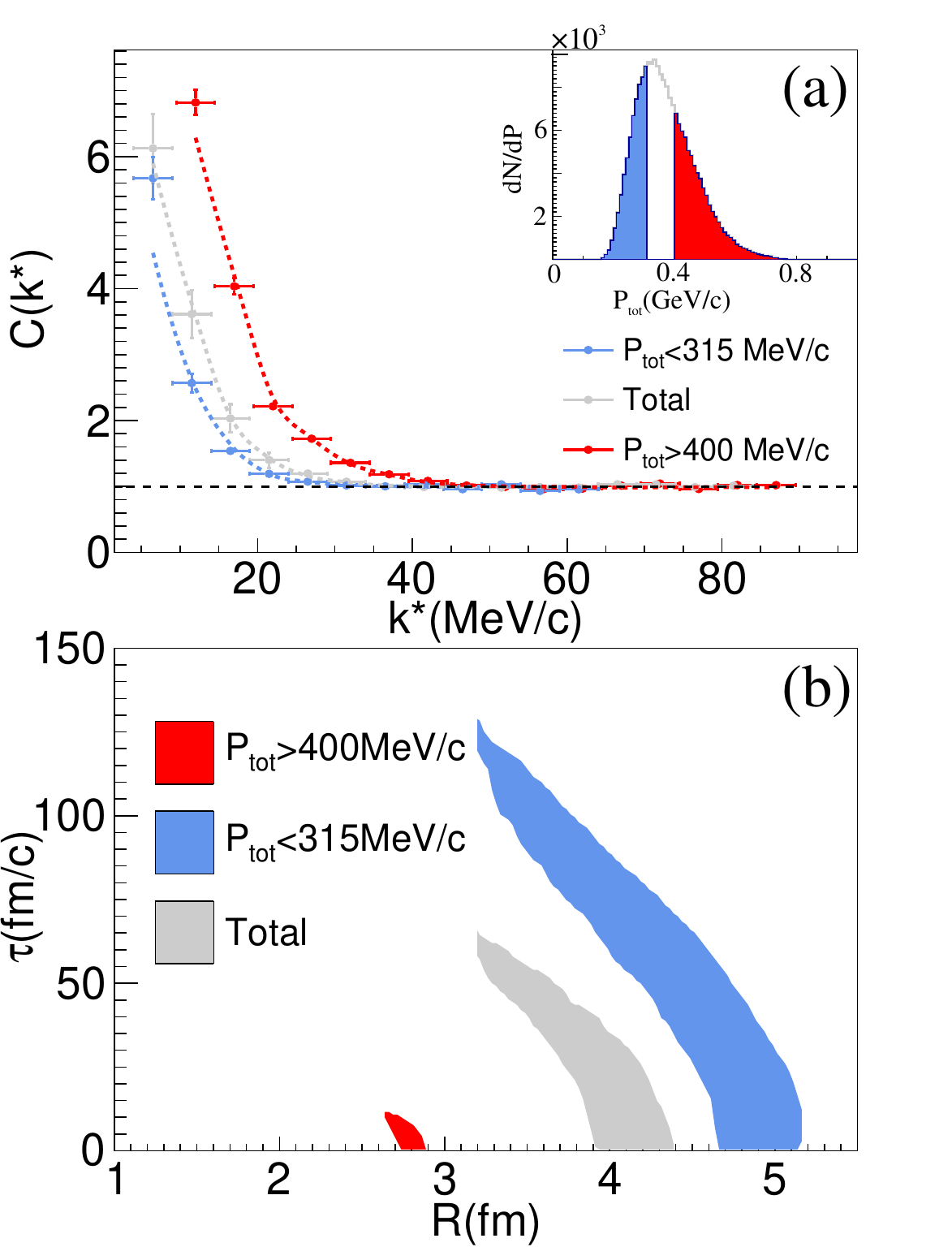} 
\caption{(Color online) (a) the corrected $nn$ CFs with different pair total momentum gates, the dashed curves represent the fittings by LL methods, the inset depicts the distribution of the total momentum of particle pairs in the laboratory frame, as well as the selected intervals corresponding to the two momentum-gated CFs. (b) the ($R$, $\tau$) contours at  99\% confidence level after LL fitting by fixing ($f_{0}^{nn}, d_0^{nn}$) for the corresponding momentum-gated CFs.}
\label{fig5}
\end{figure}

Despite of the difficulty to  simultaneously extract the emission time scale and source size through angle-averaged CF, the total momentum gated CF is a complementary way to study the interplay between dynamical and statistical effects in the particle emission \cite{nn_cor3,evolution1,PhysRevC.99.054626}. Fig. \ref{fig5} (a) presents the $nn$ CF with  high (red) and low (blue) total momentum gate in comparison with the  ungated one (gray). Enhancement (suppression) at low $k^*$ is evidently shown in the high (low) momentum-gated CFs. In order to focus on the space-time variation with the pair momentum, we fix physically $f_{0}^{nn}=18.9~ \rm fm $ and $d_{0}^{nn}=1.9~ \rm fm$, but keep $R$ and $\tau$  as free parameters in the fitting of the momentum-gated CFs. The contour plots of 99\% confidence level are shown in Fig. \ref{fig5} (b). Significant momentum dependence is observed.  For the high-momentum gated CF corresponding to the early dynamic stage, the space-time ambiguity is well confined in a small space-time volume with short time constant $\tau$, while for the low-momentum gated CF, the neutrons pairs experience larger space-time distance.  Fitting the momentum gated CFs, we obtained the reduced space-time sizes of $R^{(0)}=\RHigh~ \rm fm$ and $\RLow ~\rm fm$, corresponding to the early and the late stages of emission,  respectively. This phenomenon is consistent with the experimental observations in the momentum-gated $pp$ CF \cite{pp-gated}. Given the analysis framework is well established for $pp$ CF, see for instance in \cite{Wang:2021mrv}, the upcoming combined analysis on $nn$ and $pp$ CFs in the same system will offer the opportunity to unravel the difference of the emission timescale between neutron and proton, providing novel insights into the effect of nuclear symmetry energy.  

{\it  Summary -} The $nn$ CF in $^{124}$Sn+$^{124}$Sn reactions at 25 MeV/u has been measured with CSHINE. The TOF difference between two fired neutron detectors is reproduced by using Geant4 simulations to fit the cross talk effect. For the first time, the $nn$ scattering length and effective range ($f_{0}^{nn}$, $d_{0}^{nn}$), as well as  the space-time size $R^{(0)}$  of neutron emission source are simultaneously extracted  using the  Lednick\'y-Lyuboshitz  approach in the neutron-rich heavy ion reactions. The results of ($f_{0}^{nn}$, $d_{0}^{nn}$) are consistent with the scattering experiment, which proves the validity of the LL method in describing the $nn$ CF, and demonstrates that fine femtoscopic technique in heavy-ion experiments can be facilitated  to study $nn$ interactions and  shed light on the CSB of nuclear force by reducing the uncertainty in future measurements. In addition, we have respectively observed the space-time ambiguity  by the momentum averaged CF and the evolution of the space-time size of the neutron emission source  by the momentum gated CFs. This provides new perspectives for understanding the isospin dynamics in heavy-ion reactions.

\section* {Acknowledgements}
This work is supported by the Ministry of Science and Technology under Grant Nos. 2022YFE0103400 and 2020YFE0202001, by the National Natural Science Foundation of China under Grant Nos. 12335008, 12205160 and by the Center for High Performance Computing and Initiative Scientific Research Program in Tsinghua University. The authors thank Prof. Nu Xu and Bao-An Li for their valuable discussions.


\bibliography{reference}

\end{document}